\documentclass[fleqn,10pt]{wlscirep}
\usepackage{multirow}
\usepackage[export]{adjustbox}
\usepackage[utf8]{inputenc}
\usepackage{wrapfig}
\usepackage[T1]{fontenc}
\usepackage{comment}
\usepackage{float}
\title{Two-dimensional imaging of electromagnetic fields via light sheet fluorescence imaging with Rydberg atoms}
\author[2,+, *]{Noah~Schlossberger}
\author[1,2,+]{Tate~McDonald}
\author[4]{Kevin~Su}
\author[2,3]{Rajavardhan~Talashila}
\author[4]{Robert~Behary}
\author[1,2]{Charles~L.~Patrick}
\author[1,2]{Daniel Hammerland}
\author[4]{Eugeniy~E.~Mikhailov}
\author[4]{Seth~Aubin}
\author[4]{Irina~Novikova}
\usepackage{multicol}
\author[2]{Christopher~L.~Holloway}
\author[2, $\dagger$]{Nikunjkumar~Prajapati}
\affil[1]{Department of Physics, University of Colorado, Boulder, Colorado 80309, USA}
\affil[2]{National Institute of Standards and Technology, Boulder, Colorado 80305, USA}
\affil[3]{Department of Electrical Engineering, University of Colorado, Boulder, Colorado 80309, USA}
\affil[4]{Department of Physics, College of William \& Mary, Williamsburg, Virginia 23187, USA}
\affil[+]{these authors contributed equally to this work}
\affil[*]{noah.schlossberger@nist.gov}
\affil[$\dagger$]{nikunjkumar.prajapati@nist.gov}


\begin{abstract}
The ability to image electromagnetic fields holds key scientific and industrial applications, including electromagnetic compatibility, diagnostics of high-frequency devices, and experimental scientific work involving field interactions. Generally electric and magnetic  field measurements require conductive elements which significantly distort the field.  However, electromagnetic fields can be measured without altering the field via the shift they induce on Rydberg states of alkali atoms in atomic vapor, which are highly sensitive to electric fields. Previous field measurements using Rydberg atoms utilized electromagnetically induced transparency to read out the shift on the states induced by the fields, but did not provide spatial resolution. In this work, we demonstrate that electromagnetically induced transparency can be spatially resolved by imaging the fluorescence of the atoms. We demonstrate that this can be used to image $\sim$ V/cm scale electric fields in the DC-GHz range and $\sim$ mT scale static magnetic fields, with minimal distortion to the fields. We also demonstrate the ability to image $\sim$ 5~mV/cm scale fields for resonant microwave radiation and measure standing waves generated by the partial reflection of the vapor cell walls in this regime. 
With additional processing techniques like lock-in detection, we predict that our sensitivities could reach down to nV/cm levels. 
We perform this field imaging with a spatial resolution of 160 $\mu$m, limited by our imaging system, and estimate the fundamental resolution limitation to be 5 $\mu$m. 
\end{abstract}

\begin{document}

\flushbottom
\maketitle

\thispagestyle{empty}
\section*{Introduction}

\begin{figure*}
\centering

\includegraphics[scale = 0.9]{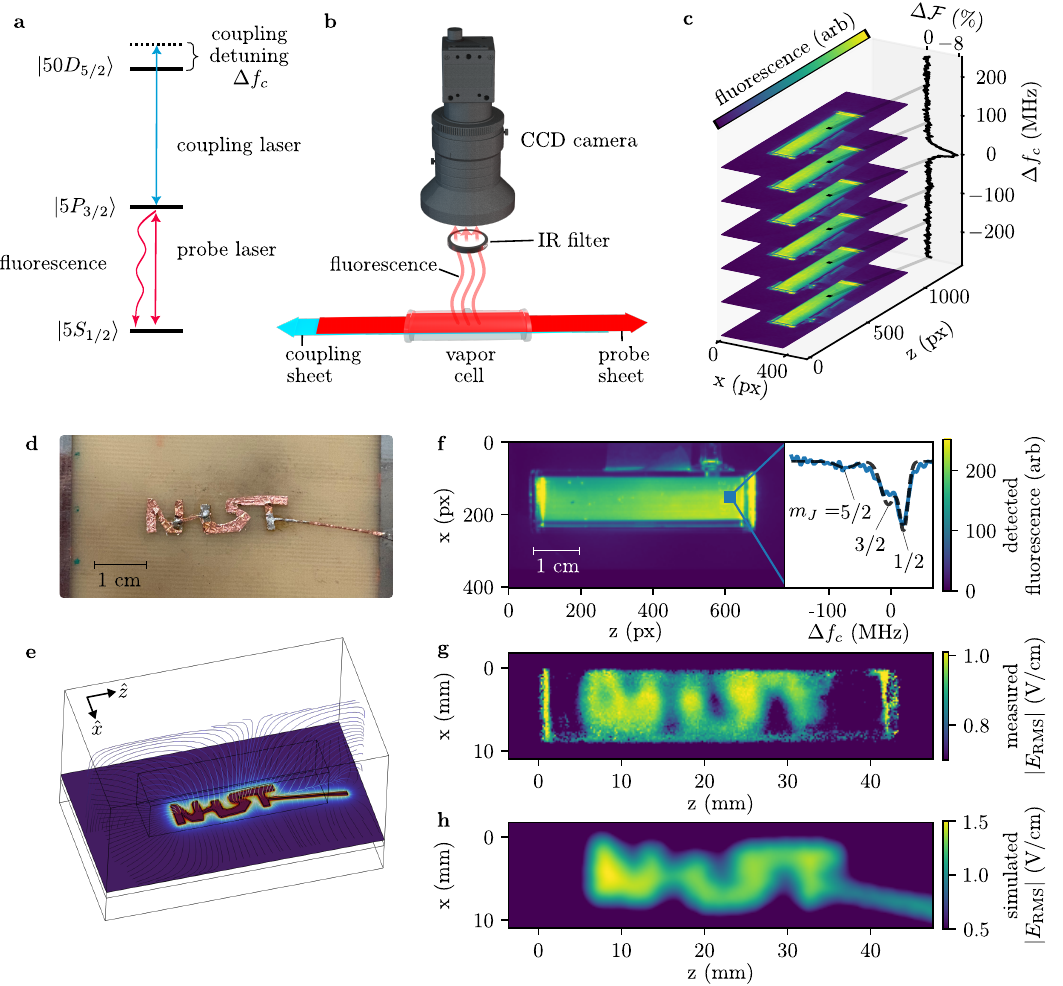}\\
\caption{\textbf{The fluorescence measurement scheme.}   \textbf{a}, The energy level diagram for $^{85}$Rb used in the measurement. \textbf{b}, The experimental setup for spatially resolved fluorescence measurement. \textbf{c}, The CCD image of the light sheet is recorded as the coupling laser is scanned ($\Delta f_c$). The resulting change in fluorescence $\Delta \mathcal{F}$ of a 2$\times$2 pixel average in the vapor cell (representing a 160$\times$160~$\mu$m area) is shown in black.
\textbf{d},~The NIST logo is constructed of copper over a 1.6 mm thick circuit board with a conductive ground plane underneath. 
\textbf{e},~The geometry is replicated in a finite element model. 
\textbf{f},~The fluorescence is imaged in an 11$\times$11$\times$45~mm cell placed above the letters, with the beam placed 2.5~mm above the letters. 1~V$_\textrm{pk}$ at 30~MHz is applied to the letters. The inset shows the spectrum taken from a group of (2$\times$2) that were averaged to achieve sufficient signal to noise. The electric field strength is found by fitting the spectrum to the Eq. \ref{eq:ACStarkFit}. 
\textbf{g},~The field is measured at all points in the cell by fitting each point. \textbf{h}, The simulated field strength at 2.5~mm above the letters given by the finite element model.}
\label{fig:IntroFig}
\end{figure*}

Imaging of electric and magnetic fields has a diverse range of applications that extend across many disciplines. 
Field imaging is an essential tool in fields ranging from biomedicine to communications, defense, electronics, and scientific research. 
In communications and defense, spatial resolution of electric fields plays a crucial role in range finding, radar\cite{moreira2013tutorial}, angle-of-arrival estimation\cite{AOA}, and beam-forming\cite{beam_form}. For integrated-circuits, field imaging can be used in quality control and performance evaluation. In electromagnetic compatibility testing, emissions of devices must meet stringent regulations\cite{paul2022introduction}. Field imaging is useful both to evaluate devices on their emissions as well as to understand the modes of radiation leakage and inform the design of these devices. These diverse applications rely on a variety of sensors to achieve precise and sensitive measurements of electric and magnetic fields. For example, microwave and radio frequency (RF) radiation is typically measured using electrically conductive antennas, each designed to operate effectively within specific frequency bands\cite{Pozar:882338}. The design and form factor of the receiver can vary significantly depending on the range of frequencies it needs to detect. Loop antennas are commonly used for field measurements above integrated circuits~\cite{loop_ant}. Spatial information is gained either by physically moving the antenna or by measuring the field at multiple points in space with an array of antennas. However, these conductive antennas have a crucial drawback—they tend to alter the field they intend to measure. 
Furthermore, while these point-based measurement techniques are valuable for capturing localized field data, scaling them into arrays that provide spatial information scales the cost and  technical complexity of these devices and makes them less transparent to the fields they aim to measure. 

In this manuscript, we demonstrate the capability to image both electric and magnetic fields using a Rydberg atom based sensor. Atomic sensors offer several unique advantages over conducting antennas and field probes\cite{Schlossberger2024Nature}. Because they consist of dielectric materials and atomic vapor, they do not tend to alter electric fields in the same way that conductors do, and do not absorb appreciable power. In addition, the same sensor geometry can be used to measure fields from DC to GHz, which would require many different sizes of antennas. Finally, atomic sensors do not require calibration to an external electric or magnetic field standard. 

These sensors measure the effect of the electric field on Rydberg states of alkali atoms in atomic vapor. In these highly excited Rydberg states, a valence electron is excited to order 1~$\mu$m away from the core, and becomes extremely sensitive to electric fields, meaning the fields can be sensitively determined by their perturbation to the spectrum of the Rydberg states. These spectroscopic measurements directly relate the field strength to a frequency shift via atomic properties and SI constants, meaning these electric field measurements are SI-traceable \cite{6910267}. While the field of Rydberg electrometry began over twenty years ago\cite{PhysRevLett.82.1831}, it does not have many results of spatially resolved measurements. Imaging of a narrow band of THz radiation resonant with a Rydberg-Rydberg transition has been performed in one dimension via its induced Autler-Townes splitting\cite{Wade2017} and in two dimensions via its induced population transfer \cite{PhysRevX.10.011027}. However, these techniques only work for a discrete set of frequencies of THz radiation. In addition, readout of the Rydberg states in a vapor cell with electromagnetically induced transparency has been used to image fields in one dimension by scanning the position of the laser beams\cite{10.1063/1.4883635,Fan:14,Ma:20}. This measurement is not truly spatially resolved: the measured spectrum represents an integral of the spectra due to the fields along the length of the beam. Because of this, it is only useful for measuring field distributions that are uniform in one dimension. Localizing the measurement by crossing the laser beams comes at the expense of severe deterioration of spectral resolution\cite{Su2024}.

In contrast, we propose a technique that works over a broad, continuous range of frequencies and is truly spatially resolved. Typical electric field measurements using Rydberg atoms follow the work in Ref.\cite{Sedlacek2012}, measuring the energy of the Rydberg states using two-photon spectroscopy. Normally, the transmission of the one photon is monitored as the other is scanned through resonance. However, the resonance condition can also be detected via the change in fluorescence \cite{PhysRevLett.81.4592, 10.1116/5.0201928}. Because the fluorescence is spatially resolved, spectroscopy can be performed independently at different points in space by imaging the fluorescence. To achieve two-dimensional imaging, the two laser fields are expanded into light sheets, and the fluorescence in this plane is imaged with a camera. Electromagnetic fields can then be imaged by the shifts they induce on the spectrum, which are easily calculable with industry-standard code libraries\cite{ROBERTSON2021107814, MILLER2024108952}.  This technique can be used to measure both electric fields via the Stark shift they induce or magnetic fields via the Zeeman shift they induce. This work represents a truly spatially localized measurement of broadband electromagnetic fields, whereby the same configuration can be used to measure electric fields continuously from DC to GHz, as well as static magnetic fields.

\subsection*{Measurement scheme}

The measurement scheme is shown in Fig. \ref{fig:IntroFig}. Ground state atoms are coupled to a Rydberg state via a two-photon ladder scheme (Fig. \ref{fig:IntroFig}a). The ground state is coupled to an intermediate state with a 780~nm probe laser, and the intermediate state is coupled to Rydberg states with a 480~nm coupling laser. When both the probe and the coupling laser are resonant, the population will enter a dark state and the probe absorption (and thus the fluorescence of the intermediate state) will decrease. This effect is called electromagnetically induced transparency (EIT)\cite{PhysRevLett.98.113003}. As the detuning  ($\Delta f_c$) of the coupling laser is scanned, the fluorescence will dip when the coupling laser is on resonance, thus reading out the energy of the Rydberg state. The experimental setup is shown in Fig. \ref{fig:IntroFig}b. The probe and coupling beams are formed into light sheets with anamorphic prism pairs and cylindrical lenses. They are sent into the vapor cell counter-propagating such that the Doppler shift due to thermal velocity of the atoms is partially canceled. The fluorescence then passes through a 780$\pm$10~nm optical bandpass filter and is imaged by a CCD camera with a compound lens focused on the plane of the beams. The resulting image is shown in Fig. \ref{fig:IntroFig}c.  The coupling laser's detuning $\Delta f_c$ is scanned with a period of  10~S, and the CCD camera records the fluorescence at a framerate of 100~Hz, so the time axis gives a spectrum at each point on the image. Both the $50D_{5/2}$ and the nearby $50D_{3/2}$ state can be seen on the spectrum, and the known energy separation of these states (92.7 MHz) allows for the calibration of $\Delta f_c$ from the frame number. The fluorescence spectrum is measured to have a Gaussian one-sigma width of 9.9$\pm$0.2~MHz.

As an example, a measurement of the AC electric field is shown in Figure \ref{fig:IntroFig} d-h. An oscillating voltage with an amplitude of 1~V and a frequency of 30~MHz is applied to a conducting sheet which spells out the word ``NIST'' (Fig. \ref{fig:IntroFig}d). The electrode is placed over a 1.6~mm thick sheet of dielectric with a ground plane underneath. A Rb vapor cell is then placed over the logo, and the light sheet is placed just above the glass wall of the vapor cell, about 2.5~mm above the electrodes. We then imaging fluorescence from the light sheet to perform spectroscopy of the vapor of atoms at each point in space (Fig. \ref{fig:IntroFig}f). The spectrum at each point in space is then fit to find the magnitude of the field (Fig. \ref{fig:IntroFig}g). A finite element model of the electrode geometry (Fig. \ref{fig:IntroFig}e) provides a theoretical field distribution in the plane of the image (Fig. \ref{fig:IntroFig}h), which matches the measured field distribution. To fit the spectra, we normalize the measured fluorescence spectra and minimize the residuals from Eq. \ref{eq:ACStarkFit} with a common set of weights for all locations, adjusting only the electric field at each point.

Infrared-absorbing matte black paper was placed beneath the cell to prevent reflections of the fluorescence from surfaces beneath the cell back to the camera. The probe and coupling powers incident on the cell are 610 $\mu$W and 380 mW respectively, and are shaped into light sheets that are 12~mm wide with a flat top profile in $\hat x$ and 1~mm at full width half maximum with a Gaussian profile in $\hat z$. The vapor cell is left at room temperature (293~K). The images are taken through a 25~mm diameter spectral band-pass filter at a distance of 15~cm (with the camera accepting light over a broader angular range), resulting in an effective numerical aperture of 0.08 and a collection efficiency of 0.2\%. The projected pixel size is calibrated using the size of the vapor cell.
\section*{Results and discussion}
\subsection*{Non-resonant electric fields}

Non-resonant electric fields can be measured via their induced Stark shift on the Rydberg state. Generally, each angular momentum substate $m_J$ will undergo a different shift, causing the spectrum to split into different peaks. The frequency shift on each $m_J$ level due to the Stark shift for weak electric fields is given by
\begin{equation}
    \Delta f_\textrm{Stark}({m_J},E) \approx -\frac{1}{2}\alpha_{m_J} E^2/h,
\end{equation}
where $\alpha_{m_J}$ is the polarizability of the state, $h$ is Planck's constant, and $E$ is the root-mean-square (RMS) value of the electric field. In the DC to 5 GHz regime, the polarizabilities $\alpha_{m_J}$ are relatively frequency independent (see Supplemental information), meaning we can treat the effect at these frequencies as a DC Stark shift using the RMS value of the field.

To detect electric fields via their Stark shift, they must shift the states by more than one linewidth, typically requiring fields on the order of V/cm. In this regime, the response is no longer quadratic, as avoided crossings begin to repel nearby states. This can even cause the shifts to change sign at certain field strengths. To account for this, we numerically calculate a Stark Map using the ARC python package\cite{ROBERTSON2021107814} (see Supplemental information) to find the Stark shifts $\Delta f_\textrm{Stark} (m_J, E)$. We can then fit our spectrum to a sum of Gaussians at the measured EIT linewidth $\sigma$ with an emperical set of weights $A_{m_J}$ in order to extract the field:
\begin{equation}
  \Delta \mathcal{F}_{\textrm{Stark fit}} = \sum_{m_J} A_{m_J}\exp\left(-\frac{(\Delta f_c-\Delta f_\mathrm{Stark} (m_J, E))^2}{2\sigma^2}\right), \label{eq:ACStarkFit}
\end{equation}

\begin{wrapfigure}{r}{.4\linewidth}
\centering
\includegraphics[scale= .9]{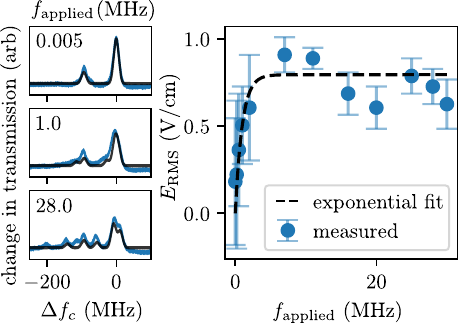}
\caption{\textbf{Shielding of low frequency electric fields by the vapor cell}. An electric field at a fixed amplitude of 0.71 V$_\mathrm{RMS}$ is applied using electrodes placed 11~mm apart on either side of the cell. The applied frequency $f_\textrm{applied}$ is then scanned, and an EIT spectrum is measured (left) at various $f_\textrm{applied}$ (blue) and fit (black) to find the electric field strength. This is repeated for many $f_\textrm{applied}$ and the rolloff at low frequencies is mapped (right). The data is fitted to an exponential fit and the cutoff frequency is found to be 1.0$\pm$0.6~MHz.}
\label{fig:Shielding}
\end{wrapfigure}

The introduction of a vapor cell inside an electric field in the MHz-GHz range is relatively unintrusive inside the MHz-GHz regime. However, near DC (<1~MHz), atomically thin layers of Rubidium can act conductively to prevent fields from entering the cell (Fig. \ref{fig:Shielding}). Making Rubidium vapor cells from monocrystalline sapphire seems to reduce the cutoff of the low frequency screening effect to around 1~kHz \cite{PhysRevApplied.13.054034}. Furthermore, DC electric fields can be imaged if they are generated using electrodes inside the vapor cell, as is done without spatial resolution in Ref. \cite{10.1116/5.0097746}. On the other hand, higher frequencies in the regime where the wavelengths are the same order as the size of the cell (10s of GHz and above) will form standing waves inside of the cell \cite{10.1063/1.4883635, PhysRevApplied.4.044015} as the cell walls become partially reflective. These effects limit the frequency range of non-disruptive imaging of external electric fields to the 10 MHz to 1 GHz regime.

\begin{figure*}[h]
\centering
\includegraphics[scale = 0.9]{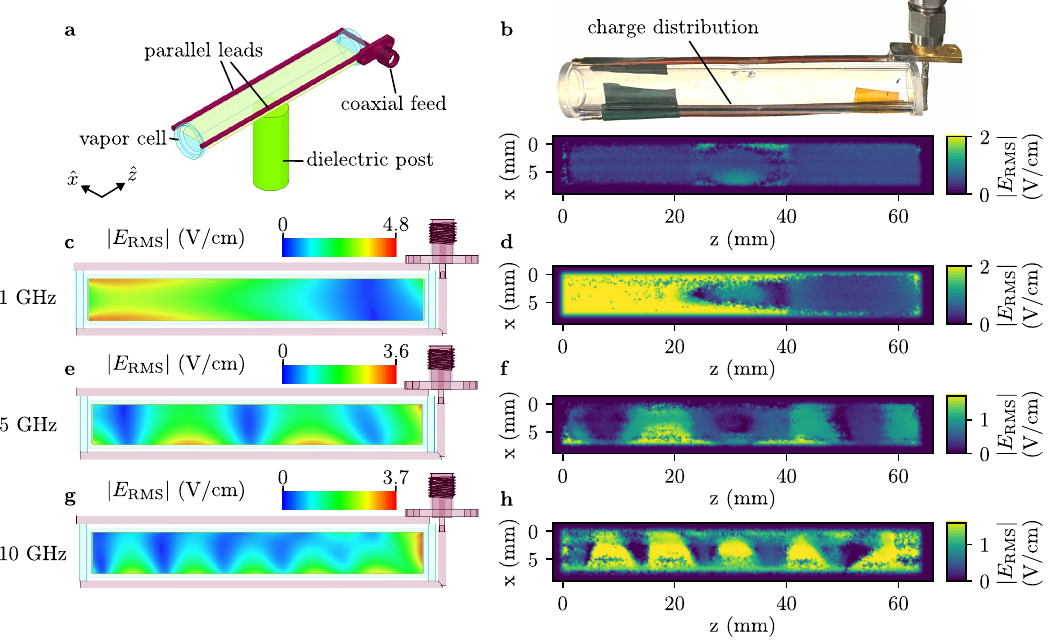}
\caption{\textbf{Fluorescence images of standing waves in a parallel wire transmission line.} The finite element model (\textbf{a}) and actual transmission line (\textbf{b}) are shown. The fluorescence imaging is performed in the plane definied by the leads. \textbf{b}, The vapor cell used in this dataset contained a patch of charges opposite the stem of the vapor cell. This is imaged to demonstrate DC field sensing capabilities inside of the vapor cell, illustrating the ability to image fields due to localized charges. \textbf{c}-\textbf{h},  A signal at 15~dBm is applied and the resulting fields are shown at 1~GHz (simulation \textbf{c}, measured \textbf{d}), 5~GHz (\textbf{e},\textbf{f}), and 10~GHz (\textbf{g},\textbf{h}).}
\label{fig:StarkTline}
\end{figure*}

To demonstrate the electric field imaging capability (in addition to Fig. \ref{fig:IntroFig}), we use a two wire transmission line with an open end. The wires are separated by 11~mm and are 75~mm long, and a 10~mm diameter $\times$ 75~mm length cylindrical vapor cell is placed inside. The measured fields are shown in Fig. \ref{fig:StarkTline}. The vapor cell we used had a patch of charge near the stem, allowing us to measure a DC electric field distribution. We then used the well-defined fields inside the waveguide to measure AC electric fields up to 10 GHz.  The reflection at the open end creates well-defined standing modes which vary along z. We measure the fields at 1 GHz where one half-wavelength fits along the transmission line, at 5 GHz where about three full wavelengths fit along the transmission line, and at 10 GHz where about six full wavelengths fit along the transmission line.  

\subsection*{Resonant microwave fields}
If a microwave field is at a frequency resonant with a Rydberg-Rydberg transition, the Rydberg states of the atom are much more sensitive to it, and therefore much weaker fields can be detected. Resonant radiation leads to Autler-Townes splitting of the Rydberg state, in which the EIT peak splits into two peaks separated by the a frequency $\Delta f_\textrm{AT}$ of:
\begin{equation}
\Delta f_\textrm{AT} = \frac{\mu_\textrm{RF} E}{h},
\end{equation}
where $E$ is the peak value of the applied resonant electric field, $h$ is Planck's constant, and $\mu_\textrm{RF}$ is the transition dipole moment of the Rydberg-Rydberg transition. We can thus determine the field by fitting the EIT spectrum to:
\begin{equation}
\Delta \mathcal{F}_\textrm{AT fit} = \exp\left({\frac{-\left(\Delta f_c - \frac{\mu_\textrm{RF} E}{2h}\right)^2}{2\sigma^2}}\right) +\exp\left({\frac{-\left(\Delta f_c + \frac{\mu_\textrm{RF} E}{2h}\right)^2}{2\sigma^2}}\right). 
\label{eq:ATfit}
\end{equation}

\begin{figure}[h]
\centering
\includegraphics[scale = 0.9]{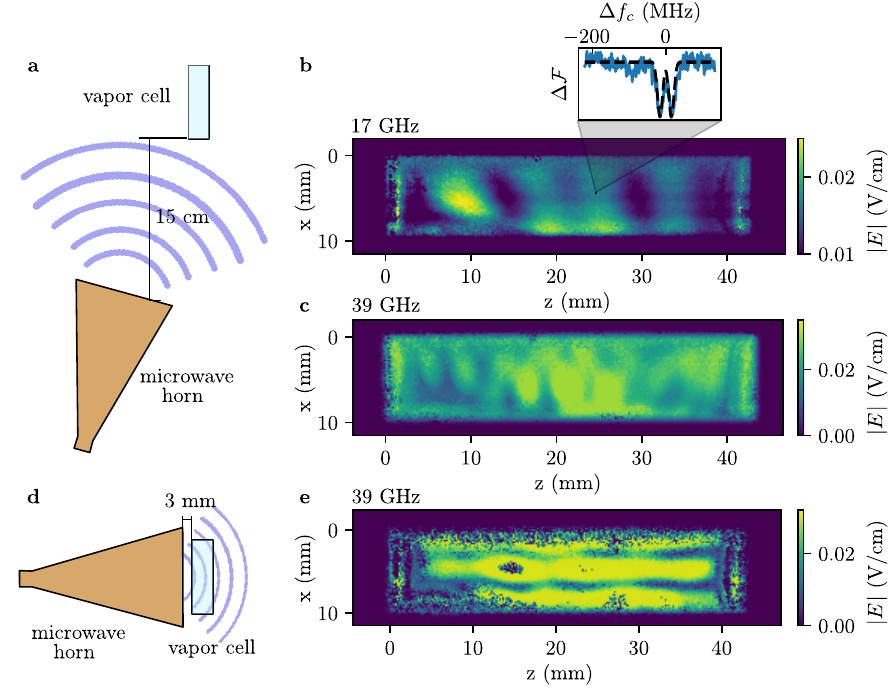}
\caption{\textbf{Imaging of standing waves of RF radiation in a vapor cell via Autler-Townes splitting.} \textbf{a}, An illustration of the setup. The horn is placed slightly off axis to allow for the optical beams. \textbf{b}, The fluorescence spectrum at each point is fitted to Eq. \ref{eq:ATfit} to extract the field magnitude. An example spectrum is shown in the inset, with the fit function overlayed (black, dashed). Here, -4~dBm of 17.041 GHz is applied to the horn.  \textbf{c}, The measurement is repeated at 39.385 GHz. \textbf{d}, Next, the horn is placed against the wall of the cell. \textbf{e}, The field is measured in this configuration at 39.385 GHz.}
\label{fig:2Dstandingwaves}
\end{figure}

Dipole moments for these transitions tend to be on the order of $2$~GHz/(V/cm), allowing fields on the order of 5~mV/cm to be detected. The drawback of this sensing mode is two-fold. First, only a discrete set of frequencies can be detected, and changing between them requires tuning the coupling laser of on the order of nm (although there are methods to continuously tune the Rydberg resonance \cite{PhysRevApplied.19.044049,10.1063/5.0086357,PhysRevA.104.032824}). Second, these transitions tend to be between 5 and 40 GHz. In this regime, the wavelength is not large compared to the geometry of the vapor cell, meaning that the vapor cell walls are partially reflected, and standing waves will form between them\cite{10.1063/1.4883635,PhysRevApplied.4.044015}. However, measuring the spatial distribution of the standing waves can be used to determine the angle of arrival of the microwave fields. These standing waves are imaged in an 11$\times$11$\times$45~mm cell at 17.041 GHz, resonant with the $50D_{5/2} \rightarrow 51P_{3/2}$ transition, and at 39.385 GHz, resonant with the $50D_{5/2} \rightarrow 52P_{3/2}$ transition in Fig. \ref{fig:2Dstandingwaves}.

\subsection*{Magnetic fields}
Strong magnetic fields can also be measured with fluorescence light sheet imaging of EIT by their induced Zeeman shift on the Rydberg states. While this method is not as sensitive as other atomic magnetometry techniques\cite{atomic_mags,kitch_mag}, it can be used with the same setup described above to measure fields on the order of mT. The shape of EIT spectra for Rydberg states in the presence of magnetic fields is described in Ref. \cite{PhysRevA.109.L021702}. Each $m_J$ pathway contributes to the EIT signal and will be resonant when the coupling is detuned by $\Delta f$ of 
\begin{equation}
    \Delta f (m_{F1}, m_{F2}, m_{J3}) = \frac{\mu_B B}{h}\big(\underbrace{g_{J_3} m_{J3} - g_{F_2} m_{F2}}_\textrm{coupling shift} + \underbrace{\frac{\lambda_\textrm{p}}{\lambda_\textrm{c}}(g_{F_2} m_{F2} - g_{F_1}m_{F1})}_\textrm{probe Doppler shift}\big) \label{eq:deltaF} \equiv \mu_{\textrm{eff}}B,
\end{equation}
where $\mu_B$ is the Bohr magneton, $B$ is the applied magnetic field, $h$ is Planck's constant, $f_p$ and $f_c$ are the probe and coupling laser frequencies, $m_{F1}$, $m_{F2}$, and $m_{J3}$ are the angular momentum projection quantum numbers for the ground, intermediate, and Rydberg states respectively, $g_{F_1}$, $g_{F_2}$ and $g_{J_3}$ are the Land\'e g-factors of each state, and $\mu_{\textrm{eff}}$ is the effective magnetic moment of the system for this pathway. The dominant pathway of the EIT signal are the stretch state transitions $m_{F1}=\pm 3 \rightarrow m_{F2}=\pm 4 \rightarrow m_{J3} = \pm 5/2$, which have a shift in our level scheme of $\mu_\textrm{eff} = \pm$20.4~MHz/mT. 

\begin{figure}[h]
\centering
\includegraphics[scale = 0.9]{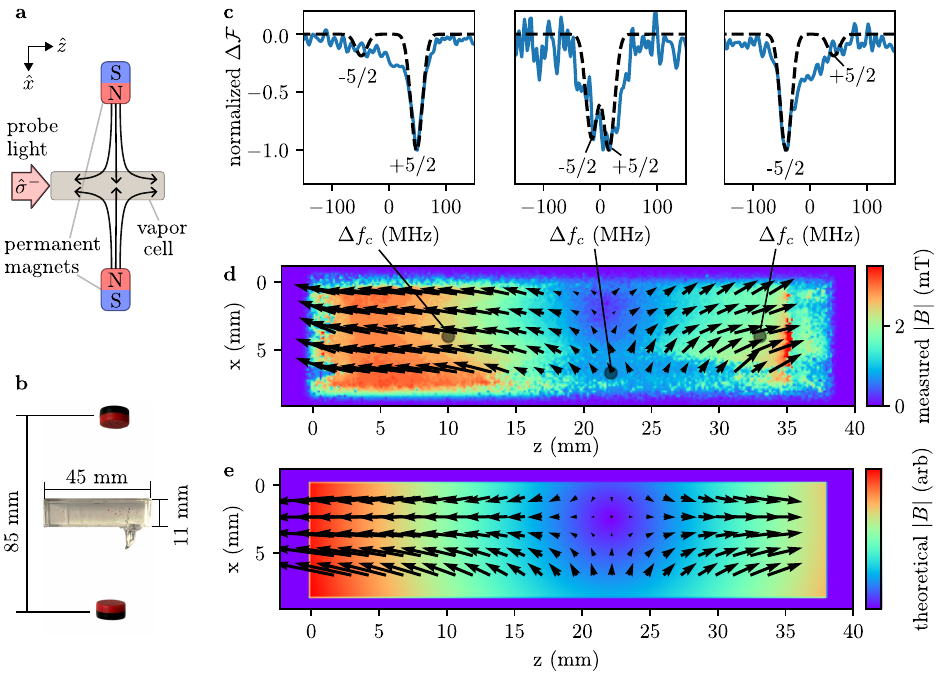}
\caption{\textbf{Fluorescence imaging of the magnetic field from two permanent magnets placed above and below the cell forming a quadrupole.} \textbf{a},  Diagram of the setup. \textbf{b}, Dimensioned image of the setup. \textbf{c}, Spectra of the fluorescence are shown for the magnetic field point left (left), upwards (center), and right (right). The fit function (black, dashed) is overlayed, and the $m_J$ level contribution to the spectra are labeled. \textbf{d}, Measured magnetic field value and vector from the fluorescence imaging fits.  Because the direction of the x-component (vertical component) of the magnetic field is ambiguous, they are plotted pointing upwards. \textbf{e}, The theoretical quadrupole magnetic field given by Eq. \ref{eq:quadTheory}.}
\label{fig:Bfield}
\end{figure}

By measuring the shift of the $m_J = \pm 5/2$ levels, we can then deduce the magnitude of the field. Furthermore, vector information can be gained by considering the distribution of $m_J$ levels. Magnetic fields at the mT level are strong enough that they set the preferred quantization axis \cite{PhysRevA.109.L021702}, meaning the optical light will appear to have different polarizations depending on this projection. To make use of this, we use circularly polarized light. Our probe light is set to $\sigma^-$ in the $z$ axis, such that if the magnetic field is in the $+z$ direction it will preferentially populate the $m_J = -5/2$ state, and the magnetic field is in the $-z$ direction it will preferentially populate the $m_J = +5/2$ state. When the magnetic field is perpendicular to the probe, the $+5/2$ and $-5/2$ states will be equally populated. Thus from the ratio of these two states on the spectrum we can get the projection of the field onto the $z$-axis independently from the magnitude of the field. These two thus give the magnitude and sign of the $z$ component of the field, and the magnitude of the orthogonal component of the field (in 2D, this is the $x$-component). Note that the sign of the $x$-component cannot be determined from this measurement.

To demonstrate this capability, we placed two permanent bar magnets 85~mm apart with their magnetic north poles facing each other in the x-axis such that they generate a quadrupole field. We then performed fluorescence imaging with an 11$\times$11$\times$45~mm rectangular vapor cell in the center of the quadrupole. The results are shown in Fig. \ref{fig:Bfield}. At each point, we fit the normalized fluorescence spectrum to a sum of two Gaussians representing the two dominant $m_J$ states:
\begin{equation}
    \Delta \mathcal{F}_{B \textrm{ fit}} = \cos^2(\theta/2) \exp\left({\frac{(\Delta f_c + \mu_\textrm{eff} B)^2}{2\sigma^2}}\right) + \sin^2(\theta/2)\exp\left({\frac{(\Delta f_c - \mu_\textrm{eff} B)^2}{2\sigma^2}}\right),
    \label{eq:Bfit}
\end{equation}
where $B$ is the applied magnetic field, $\sigma$ is the spectral width of the EIT feature = 10~MHz, and $\theta$ is the angle of the magnetic field from the z-axis.
The results (Fig. \ref{fig:Bfield}d) are consistent with the theoretical magnetic quadrupole (Fig. \ref{fig:Bfield}e), which takes the form
\begin{equation}
    B_\textrm{quadrupole} \propto (z-z_0) \hat z - (x-x_0) \hat x, \label{eq:quadTheory}
\end{equation}
where $(x_0, z_0)$ is the center of the quadrupole.

\subsection*{One-dimensional imaging}
\begin{figure*}[h]
\centering
    \includegraphics[scale = .9]{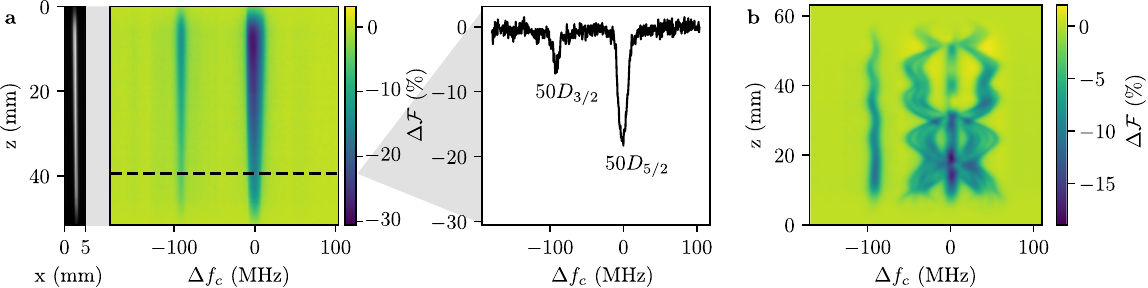}
    \caption{\textbf{One-dimensional fluorescence imaging with round beams.} \textbf{a}, An example fluorescence measurement. The fluorescence from the beam is imaged (left), and the change in fluorescence $\Delta \mathcal{F}$ is recorded in time as the coupling laser is swept (middle). The inset (right) shows the fluorescence spectrum at a single point along the beam. \textbf{b}, Standing waves are generated inside of a 25~mm diameter by 75~mm long cylindrical cell in a configuration identical to \ref{fig:2Dstandingwaves}a, and the spectrum along the beam is shown.}
    \label{fig:1Dimaging}
\end{figure*}

A main experimental limitation to this technique is the required power of the coupling laser. In a light sheet configuration, the coupling laser power is spread out, reducing the laser intensity and thus the strength of the EIT signal. Observation of strong EIT in a light sheet configuration requires coupling laser powers on the order of watts. To reduce this requirement and simplify the optics, performing fluorescence imaging in one dimension using circular beams could be advantageous. In addition, this allows for better signal-to-noise because the image can be averaged over the radius of the beam. One dimension fluorescence measurement is shown in Figure \ref{fig:1Dimaging}, demonstrating significantly better signal-to-noise for the same coupling laser power.

\subsection*{Discussion}
Imaging electromagnetic fields using light sheet fluorescence imaging of atoms coupled to Rydberg states relies on the field of interest perturbing the Rydberg spectrum. As such, the weakest field we can detect is set by the requirement that the shift induced by the field on the atomic states should be on the order of the EIT linewidth ($\sim$10~MHz) in order to resolve its effect. This limit is calculated for the three field detection mechanisms in Table \ref{tab:Fieldres}.

\begin{table}[h]
\centering
    \begin{tabular}{l|c|l}
        field detection mechanism & minimum detectable field& typical value\\
        \hline
        AC Stark (10~MHz - 1~GHz) & $\sqrt{2 h \sigma_{\textrm{EIT}} / \alpha}$ & 1 V/cm\\
        Autler-Townes (17.041 GHz) & $h \sigma_{\textrm{EIT}}/\mu_\textrm{RF}$& 5~mV/cm\\
        Zeeman (DC) & $h \sigma_{\textrm{EIT}}/\mu_\textrm{eff}$&  0.5 mT
    \end{tabular}
    \caption{\textbf{Table of electric and magnetic field sensitivities.} $h$ is Planck's constant, $\alpha$ is the largest dynamic polarizability = 204 MHz/(V/cm)$^2$, $\mu_\textrm{RF}$ is the transition dipole moment of the resonant Autler-Townes transition = 2 GHz/(V/cm), and $\mu_\textrm{eff}$ is the magnetic field response of the stretch states in the EIT spectrum = 20 MHz/mT}
    \label{tab:Fieldres}
\end{table}
In general, the field resolution can be less than these values, as our fit can resolve sub-linewidth shifts. While the amount we can split the line is dependent on the signal to noise ratio, we can generally resolve around one part in ten of the quoted minimum detectable fields.

Our current sensitivity estimation is based on measuring the shift in the atomic line. However, we have not yet utilized advanced signal processing techniques that could enhance our results. In standard transmission-based EIT experiments, which measure effects like the Stark shift or Autler-Townes splitting, fields as weak as a nanovolt per centimeter (nV/cm) have been successfully detected\cite{Jing2020}. Rather than scanning the coupling laser frequency and reading the spectrum, these methods lock the coupling laser to a particular point on the spectrum and look at amplitude changes induced by the field, which can resolve changes much smaller than a linewidth. These techniques often employ lock-in detection to improve signal-to-noise ratio. By incorporating similar signal processing methods with our camera system, we could potentially further improve the sensitivity and performance of our system. 

In our configuration, the spatial resolution was limited in practice by our imaging system to 160 $\mu$m, but this resolution can be greatly improved with a more sophisticated imaging system. The fundamental limitation of the spatial resolution comes from thermal atomic motion, with a resolution limit of on the order of 5~$\mu$m. Further discussion of spatial resolution is provided in the Supplemental information. Our imaging of magnetic and electric fields has achieved a resolution that exceeds the requirements for the fields we've generated. However, the confinement of these fields is ultimately limited by their propagation distance. The cell wall, which is over 1mm thick, plays a significant role in this limitation, as it causes the fields to lose their shape and structure. For future implimentations, we look to the fabrication of vapor cells with thinner walls for measurments with better field confinement to be used in chips characterization.

A caveat of this imaging technique is that the local fluorescence at each point is only independent in the regime where the change in transmission due to EIT is small compared to the total transmission, i.e., in the weakly absorbing regime. Otherwise, when the EIT condition is met as the laser is scanned, vapor early in the beam will transmit significantly more, causing an increase in fluorescence downstream. If this is the case, the fluorescence spectrum at any point represents the local spectrum times an integral of the transmission spectra at all points in the beam leading up to it. If the cell is heated to 315~K, we observe that the EIT signal downstream of the probe beam gets washed out and the $\Delta \mathcal{F}$ changes sign between the entrance and exit of the beam.

\section*{Conclusion}
We have demonstrated two-dimensional imaging of electric and magnetic fields with spatially resolved spectroscopy of Rydberg states of alkali atoms in a vapor cell. This technique can measure external electric fields at the V/cm level from a few MHz to a few GHz and static magnetic fields at the mT level, all with minimal distortion of the fields. We demonstrated that partial vector information can be determined from the distribution of angular momentum sub-levels in the spectrum. The technique has additional, more specialized applications in measuring resonant electric fields in the tens of GHz regime at the 5~mV/cm level, but with significant reflections from the vapor cell walls. We also show that DC electric fields generated by sources inside the vapor cell can be measured at the V/cm level. We calculate the spatial resolution to be fundamentally limited by atomic thermal velocity at 5 $\mu$m and practically limited by the resolution of our imaging system at 160 $\mu$m, allowing for deep sub-wavelength imaging of electromagnetic radiation.


\section*{Acknowledgements}
This work was partially funded by the National Institute of Standards and Technology (NIST) through the NIST-on-a-Chip (NOAC). The computational results in this work were made possible by the Baker-Jarvis high performance computer cluster in the Communications Technology Laboratory at NIST and supported by NIST's Research Services Office. The work at College of William \& Mary was supported by NSF award 2326736 and U.S. DOE contract No. DE-AC05-06OR23177.

\section*{Author contributions statement}
N.~S. conceived and conducted the experiments and analyzed results, T.~M. conducted the experiments, K.~S. conceived the experiments, R.~T. provided simulations,  C.~L.~P. conceived the experiments, D.~H. conceived the experiments, R.~B. conceived the experiments, E.~E.~M. conceived the experiments, S.~A. conceived the experiments, I.~N. conceived the experiments. C.~L.~H. secured funding and conceived the experiments, N.~P. conceived and conducted the experiments. All authors reviewed the manuscript. 

\section*{Additional information}

\textbf{Data availability} The data related to the findings of this paper are publicly available at \href{https://doi.org/10.18434/mds2-3650}{doi:10.18434/mds2-3650}.
\\
\textbf{Competing interests} The authors declare no competing interests.

\clearpage
\section*{Supplemental information}
\setcounter{figure}{0}

\begin{multicols}{2}
\subsection*{Stark shift theory}
\begin{figure}[H]
\centering
\includegraphics[]{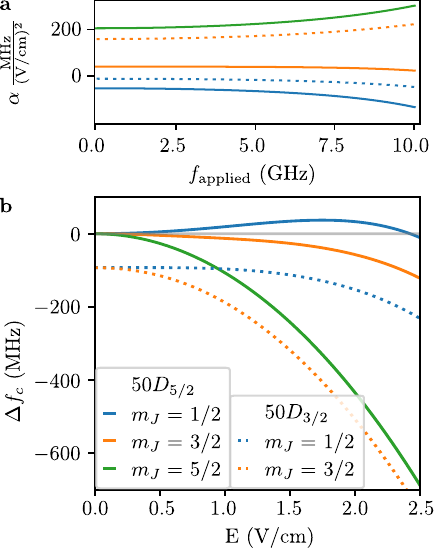}
\caption{\textbf{Response of the EIT spectrum to non-resonant electric fields.} \textbf{a}, The dynamic polarizability $\alpha$ is plotted for the relevant levels (legend in $\textbf{b}$) over a range of drive frequencies $f_\textrm{applied}$. \textbf{b}, A Stark map showing the change in the energy of each $m_J$ sub-level of the Rydberg states for various static electric field strengths.}
\label{fig:StarkMap}
\end{figure}

Here we perform calculations of the AC Stark shift due to low frequency (0-5 GHz) electric fields. We plot the polarizabilities $\alpha$ for each $m_J$ of both the $50D_{5/2}$ and $50D_{3/2}$ state in Fig. \ref{fig:StarkMap} a. The polarizabilities are calculated using the ARC python package\cite{ROBERTSON2021107814}. Because the polarizabilities are relatively frequency dependent until around 5~GHz, we can use the DC Stark map to treat these frequencies, taking the RMS value of the field. We then calculate the DC Stark map of each state using ARC. In the V/cm regime, the Stark map deviates significantly from the low-field quadratic regime. Using the full Stark map is then necessary to determine the field because the shift from each $m_J$ is not monotonic: the energy of more than one $m_J$ state is needed to constrain the magnitude of the field.

\subsection*{Resolution}

There are several physical effects that limit the spatial resolution of the effect. The main practical limit in our system is was the resolution of the imaging system, which was limited by the digital resolution of the CCD camera. The pixel size when projected onto the imaging plane was 80$\times$80 $\mu$m. To increase the signal to noise ratio, we spatially averaged every other pixel, meaning we fit spectra to find the field of 160$\times$160 $\mu$m patches. This limit could be improved by using a higher resolution imaging system to collect the fluorescence.

\begin{figure}[H]
\centering
    \includegraphics[]{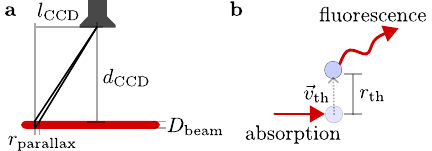}
    \caption{\textbf{Spatial resolution limiting effects.} \textbf{a}, The parallax resolution $r_\textrm{parallax}$ arises from the finite thickness of the beam resulting in a horizontal width corresponding to a vertical slice of the beam when projected onto the camera aperture. \textbf{b}, The thermal resolution $r_\textrm{th}$ arises from the atoms moving between absorbing and emitting radiation. }
    \label{fig:res_drawing}
\end{figure}

Another practical resolution limitation is parallax due to the finite thickness of the beam (Fig. \ref{fig:res_drawing} a). If the distance $d_\textrm{CCD}$ of the camera from the plane of the image is not small compared to the distance in the plane of the image to the axis of the camera $l_\textrm{CCD}$, then the finite thickness of the beam will lead to an apparent width of each portion of the beam when projected onto the camera lens. This width $r_\textrm{parallax}$ can be found by solving the triangle:
\begin{equation}
    r_\textrm{parallax} = D_\textrm{beam} \frac{l_\textrm{CCD}}{d_\textrm{CCD}},
\end{equation}
where $D_\textrm{beam}$ is the full-width at half-maximum thickness of the light sheet. In our configuration where we image a maximum of $l_\textrm{CCD}$ = 2~cm at a distance of $d_\textrm{CCD}$ = 15~cm with a light sheet of thickness $D_\textrm{beam}$ = 1.6~mm, the worst resolution due to this effect is $r_\textrm{parallax}$ = 133~$\mu$m.  Note that this effect goes to zero at the center of the image, meaning if higher resolution is needed it can be achieved by moving the camera. Additionally, this can be improved with thinner light sheets, by moving the camera farther away, or by using confocal microscopy techniques for imaging.

The previous two resolution limitations can be surpassed by changing the imaging system. However, the fundamental limit on spatial resolution is given by thermal motion of the atoms (Fig. \ref{fig:res_drawing}b). 

\vfill\null
\vspace*{.01 in}
\vspace{.5 in}

\end{multicols}\noindent
\newpage
\noindent
This motion causes the atoms to move between absorbing a probe light photon and fluorescing the photon (the dominant effect assuming that the laser fields are strong enough that atomic state adiabatically adjusts to spatial variations in Rydberg state shift). The thermal resolution is given by 
\begin{equation}
    r_\textrm{th} = \langle |\vec v_\textrm{th}|\rangle \tau,
\end{equation}
where $\tau$ is the lifetime of the intermediate state = 26 ns and $v_\textrm{th}$ is the thermal velocity, given by:
\begin{equation}
\langle |\vec v_\textrm{th}|\rangle = \sqrt{
\frac{k_B T}{m}
} ,
\end{equation}
where $k_B$ is the Boltzmann constant, $T$ is the temperature = 300~K, and $m$ is the mass of the atoms = 85~amu. In the $z$-axis, the direction of the laser beam propagation, the counter-propagating beams select a narrower velocity class of atoms that participate in electromagnetically induced transparency. In this case, the mean velocity is related to the linewidth:
\begin{equation}
\langle |\vec v_{\textrm{th},z}|\rangle = \frac{c \sigma_\textrm{EIT}}{f_p},
\end{equation}
where $c$ is the speed of light, $\sigma_\textrm{EIT}$ is the linewidth of the EIT = 10~MHz, and $f_p$ is the probe laser frequency = 384 THz. Then the x-axis thermal motion leads to a blurring of 4.5~$\mu$m and the z-axis thermal motion leads to a blurring of 0.2~$\mu$m. This effect represents a fundamental resolution limitation.
\end{document}